\documentclass[10pt]{iopart}

\usepackage{graphicx} 
\usepackage{epstopdf}
\usepackage{dcolumn}
\usepackage{bm}
\usepackage{amsfonts,amssymb
}
\usepackage{bm,dsfont}
\usepackage{bbm}
\usepackage{longtable}
\usepackage[dvips]{epsfig}
\usepackage{hyperref}
\usepackage{listings}
\usepackage{color}
\newcommand{\unit}{1\!\!1}

\begin{document}
\title{Typical bipartite steerability and generalized local quantum measurements}
\author{Maximilian Schumacher, Gernot Alber}

\address{Institut f\"{u}r Angewandte Physik, Technical University of Darmstadt, D-64289 Darmstadt, Germany}
\ead{max.schumacher@physik.tu-darmstadt.de, gernot.alber@physik.tu-darmstadt.de}

\begin{abstract}
Recently proposed correlation-matrix based sufficient conditions for bipartite steerability from Alice to Bob 
are applied to local informationally complete positive operator valued measures (POVMs) 
of the $(N,M)$-type. These POVMs allow for a unified description of a large class of local generalized measurements of current interest. It is shown that this sufficient condition exhibits a peculiar scaling property. It implies that all types of informationally complete $(N,M)$-POVMs are equally powerful in detecting bipartite steerability from Alice to Bob and, in addition, they are as powerful as local orthonormal hermitian operator bases (LOOs). In order to explore the typicality of steering numerical calculations of lower bounds on Euclidean volume ratios between steerable bipartite quantum states from Alice to Bob and all quantum states are determined with the help of a hit-and-run Monte-Carlo algorithm. These results demonstrate that with the single exception of two qubits this correlation-matrix based sufficient condition significantly underestimates these volume ratios. 
These results are also compared with 
a recently proposed method 
which reduces the determination of bipartite steerability from Alice's qubit to Bob's arbitrary dimensional quantum system to the determination of bipartite entanglement. It is demonstrated that in general this method is significantly more effective in detecting typical steerability provided entanglement detection methods are used which transcend local measurements. 
\end{abstract}
\noindent{{\it Keywords}: EPR steering,  (N,M)-POVMs,  quantum correlations, quantum information science\\}
\submitto{\PS}
\maketitle
\section{Introduction}
Basic assumptions underlying classical physics in general and special relativity in particular require consistency of all measurable statistical correlations with local realism \cite{Bell-unspeakable}. As a consequence each measurement scenario imposes particular
restrictions on measurable multipartite correlations. In general these classical correlations can be described within the framework of a local realistic theory (LRT) or local hidden variable (LHV) model \cite{Redhead}. These restrictions of possible classical correlations in local theories manifest themselves in Bell inequalities associated with classical probability polytopes \cite{Mermin,Pitowski}. A striking phenomenon of quantum systems is their capability of violating these restrictions imposed by local realism in special measurement scenarios \cite{Bell-nonlocality,Wharton}. By now this phenomenon of Bell-nonlocality has been demonstrated in a series of impressive experiments with increasing degrees of sophistication \cite{Clauser,Aspect,Zeilinger}.

The possible existence of nonlocal quantum correlations has been pointed out already as early as 1935 by Einstein, Podolsky and Rosen (EPR) \cite{EPR} in their discussion of nonlocal properties of a certain class of pure quantum states. By now the peculiar properties of these quantum states have inspired numerous applications \cite{Leuchs}. In his response to EPR Schr\"odinger termed these quantum states entangled. In addition, he coined the notion of  steering for the characteristic quantum phenomenon underlying the physical discussion of EPR, namely the ability of one local observer, say Alice, to influence simultaneously the measurements of another local observer, say Bob, by her measurements \cite{Schroedinger1,Schroedinger2}.  
But it was not until 2007 when Schr\"odinger's qualitative notion of steering, frequently also termed EPR steering, 
was clarified by Wiseman, Jones, and Doherty \cite{LHS1,LHS2}. They based their notion of EPR steering on the existence of a local hidden state (LHS) model. These LHS models generalize  LHV models by transcending the LHV-specific theoretical state concept with the help of quantum states thus allowing for a quantum theoretic description of the locally measurable probability distributions. 
This concept of EPR steerability captures some nonlocal aspects of quantum correlations and differs from Bell nonlocality and from entanglement \cite{steering-general}. Although separability of quantum states is sufficient for EPR unsteerability or stated equivalently, EPR steerability is sufficient for entanglement, the precise relation between these different concepts of nonlocal quantum correlations is intricate, in particular for mixed quantum states. Contrary to entanglement, for example, EPR steerability is an asymmetric property with respect to the local observers involved because there are multipartite quantum states which allow one local observer to steer another observer but not vice versa. Furthermore, contrary to entanglement EPR steerability is a measurement dependent notion. Whether a quantum state is EPR steerable or not depends on the local measurements which can be performed by the observers. Thus, increasing the measurement capabilities of a local observer should also increase the number of EPR steerable quantum states.

Apart from their significance for the foundations of quantum theory EPR steerable quantum states have also 
potentially promising applications in quantum information processing, in particular in the areas of quantum key distribution \cite{quantumcrypto}, quantum teleportation \cite{teleportation} and quantum secret sharing \cite{secret}. In view of this fundamental and practical significance of EPR steering
it is of interest to develop tests for EPR steerability from Alice to Bob. 
Although general necessary and sufficient conditions for bipartite EPR steerability of quantum states are known \cite{steering-general,criterion-2-qubit,Bell-diag}, these criteria involve intricate optimization procedures which even in the simplest case of two qubits require considerable numerical efforts for determining typical statistical features of bipartite EPR steerability, such as volume ratios of EPR steerable versus all quantum states, for example \cite{criterion-2-qubit}.
Up to now even in the simple case of two qubits numerically feasible necessary and sufficient conditions for EPR steerability from Alice to Bob are only known for cases in which Bob's local measurements are restricted to projective measurements. Although there is numerical evidence that these results also apply to general positive operator valued measurements (POVMs) \cite{Bell-diag} a general proof of this conjecture is still missing.

In view of these difficulties numerically less demanding conditions of EPR steerability, which are only sufficient and no longer also necessary, offer a promising alternative for investigating  typical statistical properties of EPR steerable bipartite quantum states from Alice to Bob.
In this context sufficient conditions are of particular interest which are based on measurable local correlations of Alice and Bob and which are valid for arbitrary dimensional bipartite quantum states. Recently Lai and Luo \cite{Lai} have developed such an approach based on the violation of an inequality which involves the trace norm of the correlation matrix of local measurements. As a violation of this inequality can be checked numerically in a rather straight forward way this approach may offer interesting perspectives for the exploration of statistical properties of bipartite EPR steerability from Alice to Bob.
So far, these authors have specialized their sufficient condition to cases involving local orthogonal observables, mutually unbiased measurements (MUMs) \cite{MUM} and general symmetric informationally complete positive operator-valued measures (GSIC-POVMs) \cite{GSICPOVM1}  for Alice and Bob. For these measurements they have explored bipartite EPR steerability for special families of qubit-qubit, qubit-qutrit and qutrit-qutrit states and in addition for a special one-parameter family of arbitrary dimensional bipartite isotropic quantum states.

Motivated by these developments in this paper we further explore basic properties of this recently proposed correlation-matrix based sufficient condition for EPR steerability from Alice to Bob of Lai and Luo \cite{Lai}.  In particular, we focus on two main questions. Firstly, we want to explore how the capability of detecting EPR steerability from Alice to Bob in arbitrary dimensional bipartite quantum systems depends on the nature of the local quantum measurements for a broader class of measurements. In this respect informationally complete local measurements are of particular interest, by which, at least in principle, the reduced density operators of the local observers can be reconstructed perfectly. For this purpose the recently introduced informationally complete $(N,M)$-POVMs \cite{NMPOVM} are well suited as local measurements. These families of POVMs allow for a unified description of a large class of generalized measurements of current interest, such as projective measurements involving mutually unbiased bases \cite{MUB}, mutually unbiased measurements \cite{MUM}, symmetric informationally complete generalized measurements (SIC-POVMs) \cite{SICPOVM1,SICPOVM2} and their generalized analogues (GSIC-POVMs) \cite{GSICPOVM1,GSICPOVM2}.
As a main result it will be demonstrated that the recently proposed correlation-matrix based sufficient condition for  bipartite EPR steerability from Alice to Bob of Lai and Luo \cite{Lai} exhibits a peculiar scaling property with respect to local $(N,M)$-POVMs. This scaling property implies that this sufficient condition
becomes independent of the local informationally complete measurements performed within the class of $(N,M)$-POVMs. Furthermore, it will be shown that local orthonormal hermitian operator bases (LOOs) are equally powerful to determine EPR steerability from Alice to Bob as local informationally complete $(N,M)$-POVMs. 
Secondly, we want to explore how many EPR steerable bipartite quantum states from Alice to Bob can be detected by this correlation-matrix based sufficient condition. For this purpose we sample random bipartite quantum states by a recently developed hit-and-run Monte-Carlo procedure \cite{Sauer} and determine the resulting relative Euclidean volume ratios between EPR steerable states from Alice to Bob and all possible bipartite quantum states for different dimensions of Alice's and Bob's quantum systems. This way we are capable of exploring typical statistical features of EPR steerability from Alice to Bob.  

This paper is organized as follows. For the sake of completeness in Sec. \ref{localrealism} we summarize basic facts about the basic concepts of local realism and of bipartite EPR steerability from Alice to Bob. Basic features of the recently introduced $(N,M)$-POVMs \cite{NMPOVM}, which are important for our subsequent investigation, are summarized in Sec. \ref{NMPOVM}. 
In Sec. \ref{EPR-steering} the correlation-matrix based sufficient condition for bipartite EPR steerability from Alice to Bob of Lai and Luo \cite{Lai} is introduced and applied to local measurements based on $(N,M)$-POVMs performed on arbitrary dimensional bipartite quantum systems.
In Sec. \ref{Det} it is demonstrated that within the class of local informationally complete $(N,M)$-POVMs this sufficient EPR steerability condition exhibits a characteristic scaling property. It implies that this sufficient condition becomes independent of the particular informationally complete local measurements testing for EPR steerability and, furthermore, it becomes identical with the corresponding sufficient condition for LOOs. In Sec. \ref{Numerical}
numerical results of Monte-Carlo simulations   are presented  exploring Euclidean volume ratios of steerable bipartite quantum states from Alice to Bob and all bipartite quantum states for different dimensions of Alice's and Bob's quantum systems. Furthermore, these results are compared with the corresponding volume ratios based on a
method recently proposed by Das et al. \cite{Das}, which reduces the determination of bipartite steerability from Alice to Bob to the determination of bipartite entanglement

\section{Local realism and bipartite steerability\label{localrealism}}
Classical physical theories which are consistent with the physical laws of special
 relativity and its fundamental distinction between time-like and space-like events 
 are governed by local realism \cite{Bell-unspeakable}. As a consequence classical correlations 
between two space-like separated experimenters, say Alice and Bob, obey 
locality constraints. They can be expressed in terms of generalized Bell           
 inequalities which 
 describe probability polytopes \cite{Pitowski}. 
 Basic concepts involved can be described by considering a simple bipartite scenario in which two space-like separated
 experimenters, Alice and Bob, perform random measurements of local observables, say $\alpha \in {\cal O}_A$  and $\beta \in {\cal O}_B$ with ${\cal O}_A$ and ${\cal O}_B$ describing sets of local observables of Alice and Bob. Each of these local
   observables has different possible measurement results, say $a \in {\cal M}_A$ and 
   $b \in {\cal M}_B$. 
In a local realistic classical theory (LRT) the bipartite probability distribution $P(a, b \mid \alpha , \beta)$ of a possible joint local
 measurement, in which Alice and Bob randomly select observables $\alpha$ and $\beta$ 
and obtain measurement results $a$ and $b$, has the characteristic structure
\begin{eqnarray}
P(a, b \mid \alpha , \beta) &=& \sum_{\lambda \in \Lambda} p(\lambda) 
P^A (a \mid \alpha , \lambda) P^B (b \mid \beta , \lambda),\nonumber\\
\sum_{\lambda \in \Lambda} p(\lambda)&=&1
\label{LRT}
\end{eqnarray}
with the characteristic normalization $\sum_{a \in {\cal M}_A} P^A(a \mid \alpha , \lambda) = 
\sum_{b \in {\cal M}_B} P^B (b \mid \beta , \lambda) = 1$. Here and in our subsequent discussion upper indices, such as $A$ and $B$, refer to local observers, such as Alice $(A)$ and Bob $(B)$, small Greek letters symbolize measurements and small Latin letters symbolize measurement results.
According to (\ref{LRT}), for each random selection of local observables the joint probability 
distribution is a convex sum of local (conditional) probability distributions of 
Alice and Bob, i.e. $P^A (a \mid \alpha, \lambda) \geq 0$ and 
$P^B (b \mid \beta , \lambda)\geq 0$, which depend on a random 
variable $\lambda\in \Lambda$ with probability distribution $p(\lambda)\geq 0$. 
In this description this 'hidden variable' $\lambda \in \Lambda$ 
characterizes uniquely the classical state of the bipartite physical system in a classical state space $\Lambda$.
According to Bell’s theorem \cite{Bell-unspeakable}, the constraints imposed on physical theories by local realism as expressed by Eq. (\ref{LRT}) in a bipartite scenario, for example,
can be violated by quantum correlations. In particular, the quantum correlations originating from entangled quantum state can violate local realism for particular choices of measurements. By now violations of local realism by quantum systems have been demonstrated in a series of impressive experiments with increasing degrees of sophistication
\cite{Clauser,Aspect,Zeilinger}.

Another concept which also reveals characteristic features of quantum theory is the concept of steering \cite{steering-general}. It has been introduced originally by Schr\"odinger in 1935 \cite{Schroedinger1,Schroedinger2}, in the same year in which also Einstein, Podolsky and Rosen (EPR) published their work questioning the completeness of quantum mechanical description and describing the well-known so called EPR paradoxon \cite{EPR}. Steering, sometimes also called EPR steering, may be viewed as a generalization of the experimental scenario which is described by (\ref{LRT}) within a local realistic theory and which also forms the basic scenario discussed in the EPR paradoxon. 

Bipartite scenarios are among the simplest ones by which the concept of EPR steering can be formulated. Accordingly, let us again consider two experimenters, Alice and Bob, performing general local measurements which can be described by local positive operator valued measures (POVMs) \cite{POVM-general1,POVM-general2}, say $\{\Pi^{A}_{\alpha,a}\geq 0\}$ for Alice and $\{\Pi^B_{\beta,b}\geq 0\}$ for Bob with $\alpha$ and $\beta$ denoting the different measurements and $a$ and $b$ their corresponding measurement results. These POVMs obey the characteristic completeness relations
\begin{eqnarray}
\sum_{a \in {\cal M}_A} \Pi^A_{\alpha, a} &=& {\unit}_{d_A},~~\sum_{b \in {\cal M}_B} \Pi^B_{\beta,b} = {\unit}_{d_B}
\label{POVM}
\end{eqnarray}
with the unit operators ${\unit}_{d_A}$ and ${\unit}_{d_B}$ of the $d_A$ and $d_B$ dimensional local Hilbert spaces ${\cal H}^A$ and ${\cal H}^B$ of Alice's and Bob's quantum systems. The Hilbert space of the complete bipartite quantum system is given by ${\cal H} = {\cal H}^A \otimes {\cal H}^B$.

In a typical bipartite EPR steering from Alice to Bob, Alice prepares a bipartite quantum state, say $\rho\geq 0$, with corresponding reduced local density operators  $\rho^A =\Tr_B\{\rho\}$ and $\rho^B =\Tr_A\{\rho\}$ for Alice and Bob. It is assumed that Alice can perform local measurements on the combined quantum state only with the help of her POVM $\{\Pi^{A}_{\alpha,a}\}$ and Bob can perform local measurements on this quantum state only with his POVM $\{\Pi^B_{\beta,b}\}$. 
According to quantum theory the bipartite probability distribution of a possible joint local
 measurement on this quantum state, in which Alice and Bob measure observables $\alpha$ and $\beta$ with measurement results $a$ and $b$, is given by
 \begin{eqnarray}
 P(a, b \mid \alpha, \beta,\rho) &=&\Tr_{AB}\{\Pi^A_{\alpha,a}\otimes\Pi^B_{\beta,b}\rho\}.
 \label{jointsteering}
 \end{eqnarray}
A quantum state $\rho$ is called EPR unsteerable from Alice to Bob with respect to measurements $\alpha$ and $\beta$ if this joint probability distribution can be described within the framework of a local hidden state (LHS) model \cite{LHS1,LHS2}. Otherwise this quantum state is called EPR steerable from Alice to Bob with respect to these measurements. More concretely,  EPR unsteerability from Alice to Bob with respect to local measurements described by the POVMs  $\{\Pi^{A}_{\alpha, a}\}$ and $\{\Pi^B_{\beta, b}\}$ means that
there exists a statistical ensemble of reduced quantum states of Bob, say $\{(\lambda,\rho^B_{\lambda})\mid \lambda \in \Lambda\}$ with the random variable $\{\lambda \in \Lambda\}$ being characterized by a probability distribution $\{p(\lambda)>0| \lambda \in \Lambda\}$, which is consistent with the conditional joint probability  (\ref{jointsteering}), i.e.
\begin{eqnarray}
 P(a,b \mid \alpha, \beta,\rho) &=&\sum_{\lambda  \in  \Lambda} p(\lambda) P^A(a \mid \alpha,\lambda) P^B(b \mid \beta,\lambda),\nonumber\\
 P^B(b \mid \beta,\lambda) &:=&\Tr_{B}\{\Pi^B_{\beta,b}\rho^B_{\lambda}\}.
 \label{unsteerable}
 \end{eqnarray}
 It is apparent that this restriction on the joint probability distribution of Alice $(A)$ and Bob $(B)$ is weaker than  the restriction (\ref{LRT}) imposed by local realism as it involves a quantum mechanical evaluation of Bob's local conditional probability distribution $P^B(b \mid \beta,\lambda)$.
Using (\ref{POVM}) and the characteristic normalization for conditional probabilities this condition (\ref{unsteerable}) implies the relations
\begin{eqnarray}
&&\sum_{a \in {\cal M}_A} P(a,b \mid \alpha, \beta,\rho ) =\Tr_B\{\Pi^B_{\beta, b}\rho^B\} =\\
&&\Tr_B\{\Pi^B_{\beta,b}\sum_{\lambda \in  \Lambda} p(\lambda) \rho^B_{\lambda}\},\nonumber\\
&&\sum_{b \in {\cal M}_B} P(a,b \mid \alpha, \beta, \rho ) = \Tr_A\{\Pi^A_{\alpha, a}\rho^A\} =\nonumber\\
&&\sum_{\lambda \in  \Lambda} p(\lambda) P^A(a \mid \alpha,\lambda).\nonumber
\end{eqnarray}
Thus with respect to Bob's measurement of the POVM $\{\Pi^B_{\beta,b}\}$ Bob's reduced quantum state $\rho^B$ and the LHS state $\sum_{\lambda \in  \Lambda} p(\lambda) \rho^B_{\lambda}$ are indistinguishable. 

EPR steering from Bob to Alice is defined in an analogous way. It is worth mentioning that the two concepts of EPR steerability or EPR unsteerability, namely from Alice to Bob on the one hand and from Bob to Alice on the other hand, are asymmetric. In particular, EPR steerability (or EPR unsteerability) from Alice to Bob does not necessarily imply  EPR steerability (or EPR unsteerability) from Bob to Alice. 

\section{Generalized measurements by positive operator valued $(N,M)$- measures\label{NMPOVM}}
In this section we summarize basic results on informationally complete generalized measurements which can be described by $(N,M)$-POVMs \cite{NMPOVM}. They allow for a unified description of various more specialized measurement schemes, such as projective measurements based on mutually unbiased bases \cite{MUB}, mutually unbiased measurements  (MUMs) \cite{MUM}, informationally complete generalized measurements 
(SIC-POVMs) 
 \cite{SICPOVM1,SICPOVM2}
and their generalized analogues (GSIC-POVMs) 
 \cite{GSICPOVM1,GSICPOVM2}.

In a $d$-dimensional Hilbert space a $(N,M)$-POVM, say $ \Pi$, is a set of $N$ POVMs so that each of them involves $M$ positive (semidefinite) operators describing the different possible measurement results, i.e.
$ \Pi = \{\Pi_{\alpha,a}\geq 0
|~ \alpha=1,\cdots, N; ~a =1,\cdots, M \}$. 
The parameter $\alpha$ identifies a particular measurement scheme associated with an experimental setup and the parameter $a$ identifies the corresponding different possible measurement results.
In a $(N,M)$-POVM the operators $\Pi_{\alpha,a}$ are constrained by the additional relations \cite{NMPOVM}
\begin{eqnarray}
\Tr\{\Pi_{\alpha, a} \} &=&\frac{d}{M},\label{additional1}\\
\Tr\{\Pi_{\alpha, a} ~\Pi_{\alpha, a'} \} &=&x ~\delta_{a a{'}} + (1-\delta_{a a{'}})\frac{d-Mx}{M(M-1)},\label{additional2}\\
\Tr\{\Pi_{\alpha ,a} ~\Pi_{\alpha {'}, a{'}} \} &=& \frac{d}{M^2}
\label{additional3}
\end{eqnarray}
for all $\alpha\neq \alpha {'}$.
 Different real values of  $x$ yield different $(N,M)$-POVMs,
and the possible values of $x$ are constrained by the relation
$d/M^2< x \leq {\rm min}(d^2/M^2,d/M)$.

A $(N,M)$-POVM $\Pi$ is informationally complete if and only it contains $d^2$ linearly independent positive (semidefinite) operators. As each of the $N$ POVMs fulfills the completeness relation (\ref{POVM}) informational completeness constrains the parameters $N,M$ and $d$ by the relation \cite{NMPOVM}
\begin{eqnarray}
(M-1)N + 1 &=& d^2.
\label{dimension}
\end{eqnarray} 
Therefore, each possible solution of this equation yields a possible informationally complete $(N,M)$-POVM in a given $d$-dimensional Hilbert space. As a result at
 least four possible classes of informationally complete $(N,M)$-POVMs can always be constructed. They correspond to 
 the possible solutions 
 $(N,M) \in \{(1,d^2),~(d+1,d),~(d^2-1,2),~(d-1,d+2)\}$.
In particular, the solution $(N,M) = (1,d^2)$ characterizes a one-parameter family of
generalized symmetric informationally complete positive operator valued measures (GSIC-POVM) \cite{GSICPOVM1,GSICPOVM2}.
The solution $(N,M) = (d+1,d)$ describes mutually unbiased
measurements (MUMs) \cite{MUM}, which in the special case $x=d^2/M^2=d/M=1$
 reduce to projective measurements with maximal sets of $d+1$ mutually unbiased bases.
For a qubit, i.e. $d=2$,
these four possible solutions of (\ref{dimension}) collapse to two cases, namely GSIC-POVMs ($(N,M) = (1,4)$)
and MUMs ($(N,M) = 3,2)$).
 
 In a $d$-dimensional Hilbert space
informationally complete
$(N,M)$-POVMs can be constructed with the help of a basis of $d^2$ linearly independent hermitian operators. For this purpose it is advantageous to choose an orthonormal basis of hermitian operators, i.e. $\{G_i \mid G_i=G_i^{\dagger}, i=1,\cdots,d^2\}$, with respect to the Hilbert-Schmidt (HS) scalar product $\langle G_i|G_{i'}\rangle := \Tr\{G_i^{\dagger} G_{i'}\}$. Such a basis spans the $d^2$-dimensional HS-Hilbert space ${\cal H}_{d^2}$ of hermitian linear operators over the field of real numbers and each linear operators of ${\cal H}_{d^2}$ acts on elements of ${\cal H}_d$.

Starting from an arbitrary orthonormal basis ${\cal B} = \{|k\rangle; k=1,\cdots,d\}$ in the $d$-dimensional Hilbert space ${\cal H}_d = ({\rm Span}({\cal B}), \langle \cdot |\cdot \rangle)$ of a $d$-dimensional quantum system, an example of such a hermitian orthonormal operator basis is given by
\begin{eqnarray}
\tilde{G}_1 &=& \frac{1}{\sqrt{d}}\sum_{k=1}^d |k\rangle \langle k|,\label{ONbasis}\\
\tilde{G}_{i} &=& \frac{1}{\sqrt{i(i-1)}}\left(\sum_{k=1}^{i-1} |k\rangle \langle k| - (i-1)|i\rangle \langle i| \right),\nonumber\\
\tilde{G}_{md +n} &=&\frac{1}{\sqrt{2}}\left(|m\rangle \langle n| + |n\rangle \langle m| \right),~~1\leq m<n\leq d,\nonumber\\
\tilde{G}_{(m-1)d +n} &=&\frac{i}{\sqrt{2}}\left(|m\rangle \langle n| - |n\rangle \langle m| \right),~~1\leq n<m\leq d\nonumber
\end{eqnarray}
for $i = 2,\cdots,d$.
This basis fulfills the additional convenient requirements
$\tilde{G}_1 \sqrt{d} = \unit_d$ so that  orthonormality implies $ \Tr\{\tilde{G}_{i}\}  = 0 $.
Any other orthonormal hermitian operator basis can be constructed from this one by applying an arbitrary real-valued orthogonal transformation, say $\tilde{O}$ with $\tilde{O} \tilde{O}^T = \unit_{d^2}$, onto this operator basis.
For our subsequent discussion it is advantageous to arrange the basis operators of such an arbitrary hermitian operator basis in a $d^2$-dimensional column vector, say $G = (G_1,\cdots, G_{d^2})^T= \tilde{O}\tilde{G}$.
The hermitian basis operators of any orthonormal basis fulfill the relation
\begin{eqnarray}
 \frac{1}{d} \sum_{i=1}^{d^2} G_i^2&=& \unit_d.
  \label{relations1}
   \end{eqnarray}
 Analogously, an arbitrary $(N,M)$-POVM can be combined to a $N\times M$-dimensional row vector of positive (semidefinite) operators, i.e. $\Pi=(\Pi_1,\cdots,\Pi_{NM})$. Its expansion in terms of this basis is then given by
 $\Pi = G^T S$ with the real-valued $d^2\times NM$-matrix $S$, whose  matrix elements are constrained by the defining properties
 (\ref{additional1}), (\ref{additional2}) and (\ref{additional3}) of $(N,M)$-POVMs.

\section{Correlation-matrix based sufficient conditions for bipartite EPR  steerability from Alice to Bob\label{EPR-steering}}
In this section the general form of the recently proposed sufficient condition for bipartite EPR steerability from Alice to Bob of Lai and Luo \cite{Lai} is discussed and 
applied to local measurements of Alice and Bob which can be described by arbitrary $(N,M)$-POVMs or by LOOs. It is based on a violation of an inequality involving the trace- or $1$-norm of the correlation matrix of these local measurements of Alice and Bob and is valid
for arbitrary dimensional quantum systems.

In order to investigate EPR steerability of bipartite quantum states with respect to two arbitrary local generalized measurements of the $(N,M)$-type, say $\Pi^A$ and $\Pi^B$, performed by Alice and Bob let us consider the associated correlation matrix  $C(\Pi^A,\Pi^B\mid \rho)$ with matrix elements
\begin{eqnarray}
\left(
C(\Pi^A ,\Pi^B \mid \rho)
\right)_{ij} &=& \Tr_{AB} \{ \Pi^A (i) \otimes \Pi^B (j) \left( \rho - \rho^A \otimes \rho^B \right) \}.\nonumber\\
\label{correlationmatrix}
\end{eqnarray}
Thereby, 
$\Pi^A(i)$  denotes the $i$-th component of Alice's row vector $\Pi^A$ with the indexing $i(\alpha, a) := (\alpha-1)M_A+ a \in \{1,\cdots,N_A M_A\}$ and analogously for Bob.

If  the bipartite quantum state $\rho$ is EPR unsteerable from Alice to Bob with respect to these local measurements this correlation matrix can be rewritten in the form (cf. (\ref{unsteerable}))
\begin{eqnarray}
\left(C(\Pi^A,\Pi^B\mid \rho)\right)_{ij} &=&\frac{1}{2}\sum_{\lambda,\lambda' \in \Lambda} p(\lambda) p(\lambda') V_{i}(\lambda, \lambda') W_j (\lambda, \lambda'),\nonumber\\
V_{i(\alpha, a)}(\lambda, \lambda') &=& P(a \mid \alpha, \lambda) -  P(a \mid \alpha, \lambda'),\nonumber\\
W_{j(\beta, b)}(\lambda, \lambda') &=& \Tr_{B}\{\Pi^B_{\beta, b} \rho^B_{\lambda}\} - \Tr_{B}\{\Pi^B_{\beta, b} \rho^B_{\lambda'}\}
\label{cij}
\end{eqnarray}
with $i\in \{1,\cdots,N_A M_A\}, j\in \{1,\cdots, N_B M_B\}$. The trace- or $1$-norm of this $N_A M_A \times N_B M_B$ rectangular correlation matrix can be upper bounded with the help of triangular and Cauchy-Schwarz inequalities yielding the result
\begin{eqnarray}
&&||C(\Pi^A,\Pi^B\mid \rho)||_1 \leq \\
&& \frac{1}{2}\sum_{\lambda,\lambda'\in \Lambda} p(\lambda) p(\lambda')||V(\lambda,\lambda')||_2 ||W(\lambda,\lambda')||_2
\leq\nonumber\\
&& \frac{1}{2}
\sqrt{\sum_{\lambda,\lambda' \in \Lambda}p(\lambda) p(\lambda')||V(\lambda,\lambda')||^2_2}
\sqrt{\sum_{\lambda,\lambda' \in \Lambda}p(\lambda) p(\lambda')||W(\lambda,\lambda')||^2_2}.\nonumber
\end{eqnarray}
Using the positivity of variances 
this upper bound can be further upper bounded with the help of the inequalities
\begin{eqnarray}
&&\sum_{\lambda,\lambda' \in \Lambda}\sum_{i=1}^{N_A M_A} p(\lambda) p(\lambda') \left(
V_i (\lambda, \lambda')
\right)^2\leq \nonumber\\
&&
2\sum_{i=1}^{N_A M_A}
\left(
\Tr_{A}\{
\left(\Pi^A\right)^2 (i) \rho^A\}  - 
\left(\Tr_{A}\{\Pi^A (i) \rho^A\}\right)^2
 \right) := 2 V_>,\nonumber\\
&&\sum_{\lambda,\lambda' \in \Lambda} \sum_{j=1}^{N_B M_B} p(\lambda) p(\lambda') \left(W_j(\lambda, \lambda')\right)^2\leq\nonumber\\
&& 2\sum_{\lambda \in \Lambda}\sum_{j=1}^{N_B M_B}
\left(\Tr_{B}\{\Pi^B(j) \rho^B_{\lambda}\}\right)^2 -  
2\sum_{j=1}^{N_B M_B}
\left(\Tr_{B}\{\Pi^B(j) \rho^B \}\right)^2 \nonumber\\
&&\leq 
2\left(
\max_{\sigma^B}
\sum_{j=1}^{N_B M_B}
\left(
\Tr_{B}\{\Pi^B(j) \sigma^B\}
\right)^2 -
\right. \nonumber\\&&\left.
\sum_{j=1}^{N_B M_B}
\left(
\Tr_{B}\{\Pi^B(j) \rho^B\}
\right)^2
\right) :=  2 W_>
\label{upper1}
\end{eqnarray}
which  involve a maximization over all possible reduced quantum states of Bob $\sigma^B$.
Combining these upper bound we finally obtain the inequality
\begin{eqnarray}
||C(\Pi^A,\Pi^B\mid \rho)||_1 &\leq&
\sqrt{V_> W_>}.
\label{inequality}
\end{eqnarray}
Therefore,
provided a bipartite quantum state $\rho$ is EPR unsteerable from Alice to Bob with respect to the arbitrary $(N,M)$-POVMs $\Pi^A$ and $\Pi^B$, the $1$-norm of the associated correlation  matrix, i.e. $||C(\Pi^A, \Pi^B\mid \rho)||_1$, fulfills  inequality (\ref{inequality}).
Stated differently, a
 violation of inequality (\ref{inequality}) is a sufficient condition for bipartite EPR steerability from Alice to Bob with respect to these local generalized measurements. Consequently inequality (\ref{inequality}) can be used to obtain upper bounds on measures of EPR unsteerable bipartite states from Alice to Bob and lower bounds on measures of bipartite EPR steerable bipartite states from Alice to Bob.

It is straight forward to derive an analogous inequality for an arbitrary set of local hermitian operators, say $\alpha =(\alpha_1,\cdots,\alpha_m)$ for Alice and $\beta=(\beta_1,\cdots,\beta_n)$ for Bob
describing local measurements. The resulting inequality is given by \cite{Lai}
\begin{eqnarray}
||C(\alpha,\beta \mid \rho)||_1 &\leq&
\sqrt{V^A_> W^B_>}
\label{inequality1}
\end{eqnarray}
with
\begin{eqnarray}
V^A_> &=&
\sum_{i=1}^{m}
\left(
\Tr_{A}\{
\left(\alpha_i\right)^2  \rho^A\}  -
\left(\Tr_{A}\{\alpha_i \rho^A\}\right)^2
 \right) ,\label{upper}\\
 W^B_> &=&
\left(
\max_{\sigma^B}
\sum_{j=1}^{n}
\left(
\Tr_{B}\{\beta_j \sigma^B\}
\right)^2 -
\sum_{j=1}^{n}
\left(
\Tr_{B}\{\beta_j \rho^B\}
\right)^2
\right) \nonumber
\end{eqnarray}
which  also  involves a maximization over all possible reduced quantum states of Bob $\sigma^B$.

\section{Local detection of bipartite EPR steerability from Alice to Bob by informationally complete $(N,M)$-POVMs\label{Det}}
The natural question arises how the sufficient condition on bipartite EPR steerability from Alice to Bob based on a violation of  (\ref{inequality}) depends on the type of local measurements performed by Alice and Bob. In this section it is demonstrated that inequality (\ref{inequality}) exhibits a characteristic scaling property originating from a permutation symmetry inherent in the definition of $(N,M)$-POVMs.
It implies that all informationally complete local measurements involving $(N,M)$-POVMS and all LOOs  lead to one and the same inequality whose violation yields a sufficient condition for bipartite EPR steerability from Alice to Bob. The derivation of this scaling property is based on 
general relations between informationally complete $(N,M)$-POVMs and LOOs which have been obtained recently in an investigation on the detection of typical bipartite entanglement with the help of local measurements \cite{Schumacher} and are summarized in the appendix.

In order to explore this scaling property let us first of all investigate inequality (\ref{inequality1}) for the special case of  LOOs for Alice and Bob as described by (\ref{ONbasis}). 
Accordingly, Alice measures the hermitian basis operators $\tilde{G}^A=(\tilde{G}^A_1,\cdots,\tilde{G}^A_{d_A^2})^T$ and
Bob measures the hermitian basis operators $\tilde{G}^B=(\tilde{G}^B_1,\cdots,\tilde{G}^B_{d_B^2})^T$. 
According to (\ref{upper})
the relevant upper bounds yield the results
\begin{eqnarray}
V^A_> &=&  \sum_{i=1}^{d_A^2} \left(
\Tr_{A}\{(\tilde{G}^A_i)^2 \rho^A\} - \left(\Tr_{A}\{\tilde{G}^A_i \rho^A \} \right)^2 \right) =\nonumber\\
&& d_A - \Tr_{A}\{(\rho^A)^2\},\nonumber\\
W^B_> &=& \max_{\sigma^B}  \sum_{j=1}^{d_B^2} 
\left(\Tr_{B}\{\tilde{G}^B_j \sigma^B \} \right)^2- \sum_{j=1}^{d_B^2} \left(\Tr_{B}\{\tilde{G}^B_j \rho^B\}\right)^2 =\nonumber\\
&& 1 - \Tr_{B}\{(\rho^B)^2\}.
\end{eqnarray}
Thus,  inequality (\ref{inequality1}) reduces to the result
\begin{eqnarray}
||C(\tilde{G}^A,\tilde{G}^B\mid \rho)||^2_1 &\leq&\left(d_A - \Tr_{A}\{(\rho^A)^2\}\right)
\left(1 - \Tr_{B}\{(\rho^B)^2\}\right).\nonumber\\
\label{LOO1}
\end{eqnarray}
As the $1$-norm of a matrix is invariant under arbitrary local orthogonal transformations performed by Alice and Bob, say $\tilde{O}^A$ and $\tilde{O}^B$, inequality (\ref{LOO1}) also applies to any other
LOOs for Alice and Bob, say $G^A = \tilde{O}^A \tilde{G}^A$ and $G^B = \tilde{O}^B \tilde{G}^B$, i.e.
\begin{eqnarray}
||C(G^A,G^B\mid \rho)||^2_1 &\leq&\left(d_A - \Tr_{A}\{(\rho^A)^2\}\right)
\left(1 - \Tr_{B}\{(\rho^B)^2\}\right).\nonumber\\
\label{LOO}
\end{eqnarray}

Starting from measurements involving LOOs on Alice's and Bob's sides, let us now work out inequality (\ref{inequality})  for local informationally complete $(N,M)$-POVMs for Alice and Bob by representing these POVMs in these orthonormal hermitian operator bases. Changing the LOOs from $G^A$ and $G^B$ to two informationally complete local $(N,M)$-POVMs on Alice's and Bob's sides will affect both sides of inequality (\ref{LOO}).
So, let Alice and Bob use local informationally complete $(N,M)$-POVMs, say  $\Pi^A$ and $\Pi^B$
with basis expansions $\Pi^A = (G^A)^T S^A$ and $\Pi^B = (G^B)^T S^B$ in arbitrary local orthonormal bases $G^A$ and $G^B$.
As outlined in the appendix the defining properties of informationally complete $(N,M)$-POVMs (\ref{POVM}), (\ref{additional2}), (\ref{additional3}) and (\ref{dimension}) imply that the $1$-norms of the correlation matrices exhibit the simple scaling property 
\begin{eqnarray}
||C(\Pi^A,\Pi^B\mid \rho)||_1 &=&
\sqrt{\Gamma_A \Gamma_B}||C(G^A,G^B\mid \rho)||_1
\label{scaling}
\end{eqnarray}
with
\begin{eqnarray}
\Gamma_A &=&  \frac{x_A M_A^2 -d_A}{M_A (M_A - 1)},~
\Gamma_B = \frac{x_B M_B^2 -d_B}{M_B (M_B- 1)}.
\label{Gammas}
\end{eqnarray}

Let us now investigate the scaling properties of the right hand sides of (\ref{inequality}) and (\ref{inequality1}) involving the general upper bounds (\ref{upper1}) and (\ref{upper}) which both involve a maximization over all possible reduced quantum states of Bob $\sigma^B$. Expanding also the relevant reduced density operators of Alice and Bob, i.e. 
$\rho^A = (G^A)^T r^A$, $\rho^B = (G^B)^T r^B$, and $\sigma^B = (G^B)^T s^B$,
we find with the help of (\ref{relations1}), (\ref{S1}) and (\ref{O1}) the relations
\begin{eqnarray}
&&\sum_{i=1}^{N_AM_A} \Tr_{A}\{(\Pi^A)^2(i) \rho^A\} =\Tr_{A}\{
(G^A)^T S^A  (S^A)^T G^A \rho^A
\} = \nonumber\\
&&d_A \Gamma_A +\frac{N_A d_A/M_A - \Gamma_A}{d_A},
\nonumber\\
&&\sum_{i=1}^{N_AM_A} (\Tr_{A}\{\Pi^A(i) \rho^A\})^2 = (r^A)^T S^A (S^A)^T r^A
=\nonumber\\
&& \Gamma_A \Tr_{A}\{(\rho^A)^2\} +\frac{N_A d_A/M_A - \Gamma_A}{d_A},\nonumber\\
&&
\sum_{j=1}^{N_B M_B} (\Tr_{B}\{\Pi^B(j) \rho^B\})^2 = (r^B)^T S^B (S^B)^T r^B
=\nonumber\\
&& \Gamma_B \Tr_{B}\{(\rho^B)^2\} + \frac{N_B d_B/M_B - \Gamma_B}{d_B},\nonumber\\
&&\sum_{j=1}^{N_BM_B} \max_{\sigma^B}(\Tr_{B}\{\Pi^B(j) \sigma^B\})^2 =
 \max_{s^B} (s^B)^T S^B (S^B)^T s^B
=\nonumber\\
&& \Gamma_B +  \frac{N_B d_B/M_B - \Gamma_B}{d_B}.
\end{eqnarray}
As a result inequality (\ref{inequality}) for local informationally complete $(N,M)$-POVMs of Alice and Bob becomes
\begin{eqnarray}
&&||C(\Pi^A,\Pi^B\mid \rho)||_1 =\sqrt{\Gamma_A \Gamma_B}||C(G^A,G^B\mid \rho)||_1 \leq\nonumber\\
&& \sqrt{\Gamma_A \Gamma_B} \sqrt{\left(d_A - \Tr_{A}\{(\rho^A)^2\}\right)
\left(1 - \Tr_{B}\{(\rho^B)^2\}\right)}
\label{scaling1}
\end{eqnarray}
demonstrating that a violation of inequality (\ref{LOO}) yields the sufficient condition for bipartite EPR steerability from Alice to Bob not only for arbitrary LOOs but also for arbitrary informationally complete local $(N,M)$-POVMS.

\begin{figure}
\centering
\includegraphics[width=0.75\linewidth,height=0.5\textheight]{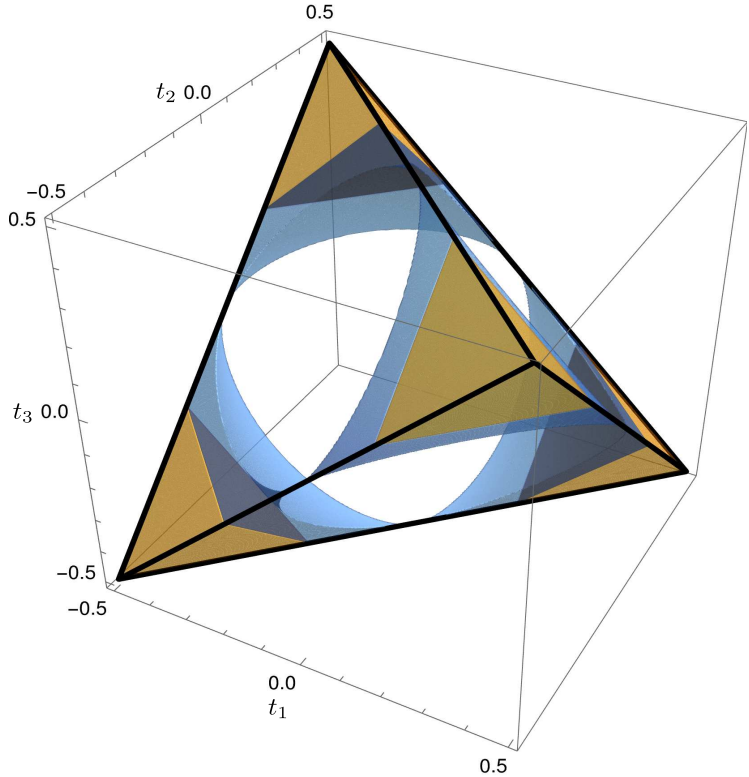}
\caption{Bell diagonal quantum states which are steerable from Alice to Bob according to a violation of inequality (\ref{scaling1}) or equivalently (\ref{LOO}) (yellow region) and according to the criterion of Uola et al. \cite{steering-general} with respect to projective measurements (yellow and blue regions): The central uncolored region represents the convex set of states which are unsteerable from Alice to Bob according to this latter criterion.
}
\label{Fig1}
\end{figure}

For the simple case of Bell diagonal states of the form
\begin{eqnarray}
\rho_{Bell} &=& \frac{1}{4}\unit_2\otimes \unit_2 + \sum_{i=1}^3 \frac{t_{i}}{2} \sigma_i^A \otimes \sigma_i^B
\label{Belldiag}
\end{eqnarray}
with the Pauli spin operators $\sigma_i^A$ and $\sigma_i^B$ for Alice and Bob it is straightforward to determine the quantum states which are steerable from Alice to Bob and which can be detected by a violation of inequality (\ref{scaling1}) or equivalently (\ref{LOO}) with the help of LOOs or of local informationally complete $(N,M)$-POVMs. In Fig. \ref{Fig1} these states are represented by the yellow regions within the tetrahedron of all Bell diagonal quantum states. The yellow and blue regions in this figure indicate the quantum states which are steerable from Alice to Bob with respect to projective measurements according to the necessary and sufficient condition of Uola et al. \cite{steering-general}. The central uncolored region is formed by the states which are unsteerable from Alice to Bob with respect to projective measurements according to this latter criterion.

The sufficient condition based on a violation of inequality (\ref{scaling1}) may be improved further by 
changing the LOOs of Alice and Bob $G^A$ and $G^B$ to another set of local hermitian operators, say $\tilde{\alpha}^A$ and $\tilde{\alpha}^B$, by relaxing the orthonormality constraints. This may be achieved by a simple rescaling of Alice's local observables, for example, i.e.
\begin{eqnarray}
	\tilde{\alpha}^A_i=h_i G^A_i
\end{eqnarray}
with the real-valued parameters $h_i $  \cite{Lai}. With these local new measurements  inequality (\ref{inequality1}) assumes the form
\begin{eqnarray}
	||C(\tilde{\alpha}^A,G^B \mid \rho)||_1 \leq&
	\sqrt{\tilde{V}^A_> \left(1 - \Tr_{B}\{(\rho^B)^2\}\right)}\label{steering_opt}
\end{eqnarray}
with
\begin{eqnarray}
	\tilde{V}^A_> &=& \sum_{i=1}^{d_A^2}h_i^2 \left(
	\Tr_{A}\{(G^A_i)^2 \rho^A\} - \left(\Tr_{A}\{G^A_i \rho^A \} \right)^2 \right).\nonumber\\
\end{eqnarray}
For a given state $\rho$ it can be tested whether the parameters $h_i$ can be chosen in such a way that inequality (\ref{steering_opt}) can be violated.  Thus, it may be possible to find for a given state $\rho$ an optimal set of parameters $h_i$ so that inequality (\ref{steering_opt}) is violated even if inequality (\ref{LOO}) is still fulfilled.
Such an optimization of the sufficient condition for EPR steerability from Alice to Bob by a variation over the parameters $h_i$ of Alice's LOO or local informationally complete $(N,M)$-POVM breaks the scaling property. Thus, a violation of inequality (\ref{steering_opt}) may be capable of detecting more EPR steerable states from Alice to Bob than a violation of inequality (\ref{LOO}).

\section{Numerical results\label{Numerical}}
Numerous investigations have already concentrated on criteria and sufficient conditions of steerability and have applied them to restricted classes of quantum states which typically form zero-measure sets within the convex set of all quantum states \cite{steering-general}. However, so far questions concerning the statistical typicality of steerability and its detectability by local measurements are largely unexplored. In this section we concentrate on this latter issues and explore Euclidean volume ratios between bipartite steerable quantum states and all bipartite quantum states. 
In particular, we concentrate on the detection of EPR steerability from Alice to Bob by local quantum measurements on the basis of  violations of inequality (\ref{steering_opt}). The required volume ratios between EPR steerable quantum states from Alice to Bob and all quantum states for different dimensions of Alice's and Bob's quantum systems will be determined with the help of 
a recently developed hit-and-run Monte-Carlo method \cite{Sauer,Sauer2}.
This way we want to shed light onto the efficiency with which typical bipartite steerability can be detected with the help of local measurements.

Starting from a $d_A\times d_B$-dimensional Hilbert space ${\cal H}_{d_A \times d_B}$ describing a bipartite quantum system the corresponding $(d_A\times d_B)^2$-dimensional Hilbert space ${\cal H}_{(d_A \times d_B)^2}$ of hermitian linear operators acting on elements of ${\cal H}_{d_A \times d_B}$ can be constructed. With respect to the Hilbert-Schmidt (HS) scalar product  $\langle A | B\rangle_{HS} := \Tr_{AB}\{A^{\dagger} B\}$ for $A,B\in {\cal H}_{(d_A \times d_B)^2}$ this HS-Hilbert space ${\cal H}_{(d_A \times d_B)^2}$ is a Euclidean vector space on which volumes of sets of hermitian linear operators and of convex sets of quantum states $\rho\geq 0$ can be defined in a natural way. Numerically the Euclidean volumes of quantum states can be determined in an efficient way with the help of a recently developed hit-and-run Monte-Carlo algorithm \cite{Sauer} which has already been applied successfully to the determination of Euclidean volumes of bipartite quantum states. This efficient Monte-Carlo method, which has been introduced originally by Smith \cite{Smith}, relies on the realization of a random walk inside a convex set that converges efficiently to a uniform distribution over this convex set and, moreover,  is independent of the starting
point inside this convex set \cite{Lovasz}.

With this hit-and-run Monte-Carlo method we have sampled $N=10^8$ bipartite quantum states randomly for different values of $d_A$ and $d_B$. For each of these randomly selected states it has been tested whether it violates the sufficient condition of EPR steerability from Alice to Bob (\ref{steering_opt})
for LOOs or informationally complete $(N,M)$-POVMs
as local measurements and for
optimized parameters $h_i$. If inequality (\ref{steering_opt}) is violated this quantum state is EPR steerable from Alice to Bob and is kept otherwise this quantum state is dismissed. This way Euclidean volume ratios $R_{S:A\to B}$ between the volumes of EPR steerable quantum states from Alice to Bob and all bipartite quantum states have been determined numerically. In view of the optimization over the parameters $h_i$ these Euclidean volume ratios $R_{S:A\to B}$ are always larger than the corresponding ratios which are obtainable directly from a violation of inequality (\ref{LOO}).

Our numerical results are summarized in Table \ref{Tab1}.
\begin{table}
  \centering
\begin{tabular}{ |c|c|c| }
\hline
Case & $d_B=2$ &  $d_B=3$  \\
\hline
$d_A=2$ & $5,011\times 10^{-2} $ &  $1,92\times 10^{-5}$ \\
&$\pm 1,5\times 10^{-4}$&$ \pm 4,1\times 10^{-6}$\\
\hline
$d_A=3$  &$5,72\times 10^{-5}$ & $0$ \\
&$ \pm 6,4\times10^{-6}$&\\
\hline
\end{tabular}
\caption{
Numerical estimates of lower bounds of  the Euclidean volume ratios $R_{S:A\to B}$ between EPR steerable quantum states from Alice to Bob and all bipartite quantum states for different dimensions $d_A$ and $d_B$ of Alice's and Bob's quantum systems: These estimates are based on a violation of inequality (\ref{steering_opt}) with Alice' local measurement being optimized by rescaling. 
The numerical errors have been estimated with the procedure described in \cite{Sauer}.}
\label{Tab1}
\end{table}
We have investigated numerically bipartite quantum states with $2\leq d_A, d_B\leq 4$. The cases not shown in Table \ref{Tab1} yield negligible volume ratios below our numerical accuracy. These results suggest that
detection of typical bipartite EPR steerability from Alice to Bob
on the basis of the sufficient condition involving a violation of inequality (\ref{steering_opt}) with LOO or informationally complete $(N,M)$-POVMs as local measurements significantly underestimates the volume ratios of EPR steerable states in higher dimensional scenarios beyond two-qubit cases. In view of the peculiar scaling properties of $(N,M)$-POVMs discussed in Sec. \ref{Det} the volume ratios of Table \ref{Tab1} cannot be increased by changes to other LOOs or informationally complete local $(N,M)$-POVMs.

In order to quantify this possible underestimation of bipartite steerability it is of interest to compare the results of Table \ref{Tab1} with the corresponding results of another sufficient condition for bipartite EPR steerability from Alice to Bob which has been proposed recently by Das et al. \cite{Das}. However, this method is only applicable whenever Alice's quantum system is a qubit. Beyond this requirement it does not impose any restrictions on the dimensionality of Bob's quantum system. These authors have proven that given such a bipartite quantum state $\rho$, entanglement of the mixed quantum state
\begin{eqnarray}
\tau_{A\to B} := \mu \rho + \frac{(1-\mu)}{2} \unit_2\otimes \Tr_A(\rho)
\label{taustate}
\end{eqnarray}
for a value of $\mu \in \left[0,1/\sqrt{3}\right]$ is sufficient for EPR steerability from Alice to Bob. As there are powerful methods for determining bipartite entanglement, this approach may possibly yield better lower bounds on the volume ratios $R_{S: A\to B}$ in its regime of validity. In particular, according to Peres \cite{Peres} and Horodecki \cite{Horodecki} one may use the existence of a negative partial transpose (NPT) of $\tau_{A\to B}$ as a sufficient condition for bipartite entanglement for cases with $d_B > 3$. In cases with $d_B \leq 3$ this latter condition is even also necessary for bipartite entanglement. However, a possible disadvantage of this approach is that it is not based on local measurements of Alice and Bob.
\begin{table}
 \hskip-1.8cm
 \centering
\begin{tabular}{|c|c|c|c|}
\hline
$d_B$ & $2$ &  $3$ & $4$\\
\hline
 & $0,05167 $ &  $0,10936$ & $0,17278$\\
 &$\pm 1,5 \times 10^{-4}$&
 $ \pm 3,4 \times 10^{-4}$&
 $\pm 5,6\times 10^{-4}$\\
\hline
$d_B$&$5$&$6$&$7$\\
\hline
&$0,24009$&$0,3119$&$0,3842$\\
&$\pm 8,3\times 10^{-4}$&
 $\pm1,3\times 10^{-3}$&
 $\pm 1,5\times 10^{-3}$\\
 \hline
\end{tabular}
\caption{
Numerical lower bounds on the Euclidean volume ratios $R_{S:A\to B}$ between EPR steerable quantum states from Alice to Bob and all bipartite quantum states for  $d_A=2$ and different dimensions $2\leq d_B\leq 7$ of Bob's quantum system: 
These estimates are based on the approach of Das et al. \cite{Das}. The Peres-Horodecki condition has been used as a sufficient condition for bipartite entanglement  of $\tau_{A\to B}$ for $d_B> 3$. The numerical errors have been estimated with the procedure described in  \cite{Sauer}.}
\label{Tab2}
\end{table}

Table \ref{Tab2} depicts our numerically obtained lower bounds on Euclidean volume ratios $R_{S:A \to B}$ based on the approach by Das et al. \cite{Das} combined with NPT tests of the states $\tau_{A\to B}$ of (\ref{taustate}) for different dimensions $d_B$ of Bob's quantum system. 
A comparison with the results of Table \ref{Tab1} demonstrates that for two qubits the detection of EPR steerability by optimized local measurements leads to a result in agreement with the one of Table \ref{Tab2}. 
However, in all other cases the local measurement based sufficient condition for EPR steerability from Alice to Bob significantly underestimates the volume ratios $R_{S:A\to B}$ of Table \ref{Tab2}. Furthermore, the volume ratios of Table \ref{Tab2} also agree with the intuition suggested by the concept of EPR steerability from Alice to Bob that increasing the dimensionality of Bob's local quantum system increases his abilities to detect EPR steerability by Alice.

In view of the differences between the results of Tables \ref{Tab1} and \ref{Tab2} one may ask whether the underestimated volume ratios of Table \ref{Tab1} may still be improved by using the approach of Das et al \cite{Das} but detecting entanglement of the quantum state $\tau_{A\to B}$ of (\ref{taustate}) by local measurements. Recently it has been demonstrated that local informationally complete $(N,M)$-POVMs are as powerful in detecting entanglement of bipartite quantum states as LOOs \cite{Schumacher}. This is a consequence of the peculiar scaling properties characterizing local informationally complete $(N,M)$-POVMs and of their relation to LOOs.
It has already been shown by Gittsovich and G\"uhne \cite{Gitt1} that a sufficient condition for bipartite entanglement detection by LOOs is given by a violation of the inequality
\begin{eqnarray}
||C(G^A,G^B|\rho)||_1 \leq 
\sqrt{\left(1 - \Tr_{A}\{(\rho^A)^2\}\right)
		\left(1 - \Tr_{B}\{(\rho^B)^2\}\right)}\nonumber\\
		\label{Gittsovich1}
\end{eqnarray}
with $G^A$ and G$^B$ denoting the LOOs of Alice and Bob and $\rho^A$ and $\rho^B$ denoting their reduced quantum states. Using LOOs of the form (\ref{ONbasis}) for Alice and Bob and the relations $\tau_{A\to B}^A = \mu \rho^A + \left((1-\mu)/2\right) \unit_2$ and $\tau_{A\to B}^B=\rho^B$ for the reduced quantum states of (\ref{taustate}), inequality (\ref{Gittsovich1}) assumes the form
\begin{eqnarray}
	\frac{1}{\mu}||C(\tilde{G}^A,\tilde{G}^B\mid \tau_{A\to B})||_1=||C(\tilde{G}^A,\tilde{G}^B\mid \rho)||_1\leq\nonumber\\
	\sqrt{\left(\frac{1+\mu^2}{2\mu^2} - \Tr_{A}\{(\rho^A)^2\}\right)
		\left(1 - \Tr_{B}\{(\rho^B)^2\}\right)}.\label{ent_correlation}
\end{eqnarray}
For $\mu=1/\sqrt{3}$ inequality (\ref{ent_correlation}) reduces to inequality (\ref{LOO}) with $d_A=2$. 
Thus, for cases with $d_A=2$ a violation of inequality (\ref{LOO}) characterizes once again the sufficient condition for EPR steerability from Alice to Bob through entanglement detection via correlation matrices of local measurements involving LOOs or informationally complete $(N,M)$-POVMs. 
This demonstrates that the approach of Das et al. \cite{Das} combined with local measurements involving LOOs or informationally complete $(N,M)$-POVMs as tests for bipartite entanglement cannot improve the  results of Table \ref{Tab1}.

\section{Conclusions}
We have applied the
correlation-matrix based sufficient condition for bipartite EPR steerability from Alice to Bob of Lai and Luo \cite{Lai} to local measurements based on $(N,M)$-POVMs performed on arbitrary dimensional bipartite quantum systems. It has been shown that
within the class of local informationally complete $(N,M)$-POVMs this sufficient EPR steerability condition, which is based on a violation of inequality (\ref{inequality}), exhibits a peculiar scaling property. It implies that a violation of one and the same inequality  characterizes this sufficient condition for measurements involving LOOs and for all informationally complete local $(N,M)$-POVMs. Thus, local informationally complete $(N,M)$-POVMs are as powerful as LOOs for detecting bipartite EPR steerability from Alice to Bob on the basis of a violation of  inequality (\ref{LOO}).

With the help of a hit-and-run
Monte-Carlo algorithm we have determined lower bounds on the Euclidean volume ratios of EPR steerable bipartite quantum states from Alice to Bob and all bipartite quantum states for low dimensions of Alice's  and Bob's  quantum systems. These numerical results explore the statistical typicality of locally detectable bipartite EPR steerability from Alice to Bob based on a violation of inequality (\ref{steering_opt}). They demonstrate that, 
except for the case of two qubits, the sufficient condition for bipartite EPR steerability from Alice to Bob resulting from a violation of inequality (\ref{steering_opt}) tends to underestimate the Euclidean volume ratios between EPR steerable bipartite quantum states from Alice to Bob and all bipartite quantum states significantly. Our numerical investigations also demonstrate that the recently introduced approach of Das et al. \cite{Das}, which relates bipartite EPR steerability from Alice's qubit to Bob's arbitrary dimensional qudit to bipartite entanglement, can be more efficient provided methods for detecting bipartite entanglement are used which transcend local measurements, such as checking for the NPT property of bipartite quantum states. However, besides not being based on local measurements a further disadvantage of this latter approach is that so far its validity is restricted to cases in which Alice's quantum system is a qubit.
Therefore, 
further research is required for the development of efficient measurement-based methods  for the detection of EPR steerability and for the exploration of its intricate relation to entanglement.

\begin{ack}
It is a pleasure to dedicate this work to Igor Jex on the occasion of his sixtieth birthday. G.A. is grateful to his friend and regular collaborator Igor Jex for numerous inspiring discussions on the intricacies of quantum physics. This research is supported by the Deutsche Forschungsgemeinschaft (DFG) -- SFB 1119 -- 236615297.
\end{ack}

\appendix
\section{}

In this appendix the derivation of the general scaling relation (\ref{scaling}) of Sec.\ref{Det} between the $1$-norms of the correlation matrices of arbitrary LOOs and local informationally complete $(N,M)$-POVMs is outlined. The general relations between informationally complete $(N,M)$-POVMs and orthonormal hermitian operators bases presented in this appendix, which this derivation is based on (cf. (\ref{S1})), have been obtained recently in an investigation on the detection of typical bipartite entanglement with the help of local measurements  \cite{Schumacher}.

We consider a $d$ dimensional Hilbert space
${\cal H}_d = ({\rm Span}({\cal B}), \langle \cdot | \cdot \rangle)$ with orthonormal basis ${\cal B}=\{|1\rangle,\cdots, |d\rangle\}$ and an associated arbitrary basis of hermitian linear operators $G = (G_1,\cdots,G_{d^2})^T$ with $G_{\mu}=G_{\mu}^{\dagger}, \mu\in \{1,\cdots,d^2\}$ acting on this Hilbert space.
 Let us also assume that this operator basis is orthonormal with respect to the Hilbert-Schmidt (HS) scalar product, i.e. $\langle G_{\mu}|G_{\nu}\rangle_{HS} := \Tr_{AB}\{G_{\mu} G_{\nu}\} = \delta_{\mu , \nu}$ so that it spans a HS-Hilbert space ${\cal H}_{d^2} = ({\rm Span}(G), \langle \cdot | \cdot \rangle_{HS})$.

An arbitrary $(N,M)$-POVM, say $\Pi = \{\Pi_1,\cdots,\Pi_{NM}\}$ with $\Pi(i)\geq 0$, $i(\alpha,a) :=(\alpha -1)M +a,~\alpha\in \{1,\cdots,N\},~a\in \{1,\cdots,M\}$, $i\in \{1,\cdots,NM\}$,
can be expanded in this orthonormal hermitian operator basis, i.e. $\Pi = G^T S$ with the $d^2\times (NM)$ matrix $S$ of real valued coefficients. Using the orthonormality of the basis $G$, for $N\geq 2$ conditions (\ref{additional2}) and (\ref{additional3}) characterizing $(N,M)$-POVMs can be rewritten in the form
\begin{eqnarray}
&&\left(S^T S\right)_{i(\alpha,a),j(\alpha',a')} =
\Gamma \delta_{i(\alpha,a),j(\alpha',a')} -\label{STS}\\
&& \frac{\Gamma}{M} \left(\bigoplus_{\alpha=1}^N J_{\alpha}\right)_{i(\alpha,a),j(\alpha',a')} +
 \frac{d}{M^2} J_{i(\alpha,a),j(\alpha',a')}\nonumber 
\end{eqnarray}
with $\Gamma = (xM^2 -d)/(M(M-1))$ (cf. (\ref{Gammas})), 
the $(NM)\times (NM)$ matrix $J$ of all ones, i.e. $J_{i(\alpha,a),j(\alpha',a')}=1$,
and with the $M\times M$ block matrices $J_{\alpha}$ of all ones, i.e.
$ \left(J_{\alpha}\right)_{i(\alpha,a),j(\alpha',a')}=\delta_{\alpha  ,\alpha'}$.
The spectrum of the positive semidefinite symmetric $(NM)\times (NM)$ matrix (\ref{STS}) is given by
\begin{eqnarray}
{\rm Sp}(S^TS) &=&\{\Gamma^{(N(M-1))},\frac{dN}{M}^{(1)}, 0^{(N-1)}\}
\end{eqnarray}
with the exponents indicating the multiplicities of the eigenvalues. For $N=1$ the zero eigenvalue no longer appears in the spectrum of $S^TS$. 
Therefore, according to (\ref{dimension}) for informationally complete $(N,M)$-POVMs the dimension $D$ of the eigenspace of the non-zero eigenvalues is given by
\begin{eqnarray}
D &=& N(M-1) +1 = d^2.
\end{eqnarray}
The spectral representation of this symmetric matrix is given by
\begin{eqnarray}
\left(S^T S\right)_{i,j} &=& \sum_{\mu=1}^{N(M-1)+1} X_{i,\mu}\Lambda_{\mu} X^T_{\mu, j}
\end{eqnarray}
 with 
 \begin{eqnarray}
 \Lambda_1 &=& \frac{d}{M^2}NM,~~X_{i,1} = \frac{1}{\sqrt{NM}},\nonumber\\
 \Lambda_{\nu} &=& \Gamma,~~\sum_{a=1}^M X_{i(\alpha,a), \nu} = 0
 \label{eigenvector}
 \end{eqnarray}
for $i\in \{1,\cdots,NM\}$, $\nu \in \{2,\cdots,N(M-1)+1\}$. 
The $(NM)\times (N(M-1)+1)$ matrix $X_{i,\mu}$ fulfills the orthogonality condition 
\begin{eqnarray}
\sum_{i=1}^{NM}\left(X^T\right)_{\mu, i} X_{i,\nu} &=& \delta_{\mu , \nu}.
\end{eqnarray} 
As a consequence of (\ref{dimension}), for an informationally complete $(N,M)$-POVM the most general form of the $d^2\times (NM)$ matrix $S$ which is consistent with (\ref{additional2}) and (\ref{additional3}) is given by
\begin{eqnarray}
S_{\mu, i} &=& \sum_{\mu'=1}^{d^2}O^T_{\mu, \mu'} \sqrt{\Lambda_{\mu'}}X^T_{\mu',i}
\label{S1}
\end{eqnarray}
with the arbitrary real-valued orthogonal $d^2\times d^2$ matrix $O$, i.e. $O O^T = O^T O = P_{d^2}$. Thereby, $P_{d^2}$ denotes the projection operator onto the $(N(M-1)+1)$-dimensional eigenspace of non-zero eigenvalues of the linear operator $S^TS$ acting in the HS-Hilbert space ${\cal H}_{NM}$. Note that in the case of an informationally complete $(N,M)$-POVM this subspace is isomorphic to the HS-Hilbert space ${\cal H}_{d^2}$.
Let us finally also add the constraint (\ref{POVM}) which yields the relation
\begin{eqnarray}
\unit_{d} &=& \sum_{a=1}^M \Pi(i(\alpha,a)) = \sqrt{d} \sum_{\mu=1}^{d^2} G_{\mu} O^T_{\mu, 1} 
\label{O1}
\end{eqnarray}
where we have taken into account the constraints (\ref{eigenvector}) on the eigenvectors of $S^TS$. This relation together with the constraints (\ref{eigenvector}) also implies condition (\ref{additional1}) so that all requirements defining an informationally complete $(N,M)$-POVM $\Pi = G^T S$, namely (\ref{POVM}), (\ref{additional1}), (\ref{additional2}), (\ref{additional3}) and (\ref{dimension}), are fulfilled.

With the help of (\ref{S1}) it is straight forward to relate the correlation matrix of two local informationally complete $(N,M)$-POVMs for Alice and Bob, say $\Pi^A$ and $\Pi^B$, to the correlation matrix of two LOOs, say $G^A$ and $G^B$.
Using relation (\ref{S1})  for these local bases we find
\begin{eqnarray}
&&C(\Pi^A,\Pi^B|\rho) = \left(S^A\right)^T C(G^A, G^B|\rho) S^B = \\
&&
X^A\sqrt{\Lambda^A} O^A C(G^A, G^B|\rho) \left( O^B\right)^T \sqrt{\Lambda^B} \left( X^B\right)^T.\nonumber
\end{eqnarray}
For the corresponding $1$-norm we obtain the result
\begin{eqnarray}
||C(\Pi^A,\Pi^B|\rho)||_1&=&||\sqrt{\Lambda^A} O^A C(G^A, G^B|\rho) \left( O^B\right)^T \sqrt{\Lambda^B}||_1\nonumber\\
\end{eqnarray}
with $\Lambda^A$ and $\Lambda^B$ denoting the diagonal matrix of nonzero eigenvalues of Alice and Bob.
Using (\ref{O1}), the degeneracy of all eigenvalues for $\mu \neq 1$, i.e. $\Lambda_{\mu \neq 1} = \Gamma$, and the invariance of the $1$-norm under orthogonal transformations
this expression simplifies to
\begin{eqnarray}
||C(\Pi^A,\Pi^B|\rho)||_1&=&\sqrt{\Gamma^A} \sqrt{\Gamma_B}||C(G^A, G^B|\rho)||_1.\nonumber\\
\label{generalscaling}
\end{eqnarray}
The above arguments demonstrate that this scaling relation is valid  for arbitrary informationally complete local $(N,M)$-POVMs $\Pi^A$ and $\Pi^B$ and LOOs $G^A$ and $G^B$.


\section*{References}
\begin{thebibliography}{63}
\bibitem{Bell-unspeakable}
Bell J S 1987 {\em Speakable and Unspeakable in Quantum Mechanics} (Cambridge University Press, Cambridge)
\bibitem{Redhead}  Redhead M 2000 {\em Incompleteness, Nonlocality and Realism} (Clarendon Press, Oxford)
\bibitem{Mermin}
Mermin N D  1983
Rev. Mod. Phys. {\bf 65} 803
\bibitem{Pitowski}
 Pitowsky I 1989 {\it Quantum Probability - Quantum Logic} Lecture
Notes in Physics {\bf 321} (Springer, Berlin)
\bibitem{Bell-nonlocality}
 Brunner N,  Cavalcanti D, Pironio S,  Scarani V and  Wehner S 2014
Rev. Mod. Phys. {\bf 86} 419
\bibitem{Wharton}
 Wharton K B and Argaman N 2020
Rev. Mod. Phys. {\bf 92} 021002
\bibitem{Clauser} 
Freedman J and Clauser J F 1972 Phys. Rev. Lett. {\bf 28}
938
\bibitem{Aspect} Aspect A, Grangier Ph and Roger G 1981
Phys. Rev. Lett. {\bf 47} 460
\bibitem{Zeilinger} 
Zeilinger A 1999
Rev. Mod. Phys. {\bf 71} S288
\bibitem{EPR}  Einstein A, Podolsky B and Rosen N 1935 Phys. Rev. {\bf 47} 777
\bibitem{Leuchs}
Reid M D,  Drummond P D,  Bowen W P,  Cavalcanti E G, Lam P K, Bachor H A,  Andersen U L and Leuchs G 2009
Rev. Mod. Phys. {\bf 81} 1727
\bibitem{Schroedinger1} Schr\"odinger E 1935 Math. Proc. of the Camb. Philos. Soc. {\bf 31} 555
\bibitem{Schroedinger2} Schr\"odinger E 1936 Math. Proc. of the Camb. Philos. Soc. {\bf 32} 446
\bibitem{LHS1} Wiseman H M, Jones S J and Doherty A C 2007 Phys. Rev. Lett. {\bf 98} 140402
\bibitem{LHS2} Jones S J,  Wiseman H M and Doherty A C 2007 Phys. Rev. A {\bf 76} 052116
\bibitem{steering-general} Uola R,  Costa A C S, Nguyen H C and G\"uhne O 2020  Rev. Mod. Phys. {\bf 92} 015001
\bibitem{quantumcrypto} Branciard C, Cavalcanti E G, Walborn S P, Scarani V and Wiseman H M 2012 Phys. Rev. A  {\bf 85} 010301 (R)
\bibitem{teleportation}  He Q, Rosales-Zarate L,  Adesso G and Ried M D 2015 Phys. Rev. Lett. {\bf 115} 180502
\bibitem{secret}  Xiang Y,  Kogias I, Adesso G and He Q 2017 Phys. Rev. A {\bf 95} 010101 (R)
\bibitem{criterion-2-qubit}  Yu B-C,  Jia Z-A,  Wu Y-C and Guo G-C 2018 Phys. Rev. A {\bf 97} 012130
\bibitem{Bell-diag} 
Nguyen H C,  Nguyen H-V and G\"uhne O 2019 Phys. Rev. Lett. {\bf 122} 240401
\bibitem{Lai} Lai L and  Luo S 2022 Phys. Rev. A  {\bf 106} 042402
\bibitem{MUM}  Kalev A and  Gour G 2014 New J. Phys. {\bf 16} 053038
\bibitem{GSICPOVM1} Gour G and Kalev A 2014 J. Phys. A: Math. Theor. {\bf 47} 335302
\bibitem{NMPOVM}  Siudzi\'{n}ska K 2022 Phys. Rev. A {\bf 105} 042209
\bibitem{MUB} Wootters W K and Fields B D 1989 Ann.Phys. {\bf 191} 363
\bibitem{SICPOVM1}  Renes J M,  Blume-Kohout R,  Scott A J and  Caves C M 2004 J. Math.Phys. {\bf 45} 2171
\bibitem{SICPOVM2} Rastegin A E 2014 Phys. Scr. {\bf 89} 085101
\bibitem{GSICPOVM2} Yoshida M and Kimuar G 2022 Phys. Rev. A {\bf 106} 022408
\bibitem{Sauer} Sauer A, Bern\'{a}d J Z,  Moreno H J and Alber G 2021
J. Phys. A: Math. Theor. {\bf 54} 495302
\bibitem{Das}  Das D,  Sasmal S and  Roy S 2019 Phys. Rev. A {\bf 99} 052109
\bibitem{POVM-general1} Holevo A S 2001, {\em Statistical Structure of Quantum Theory} (Springer, Berlin)
\bibitem{POVM-general2} Bergou J A, Hillery M S and  Saffman M 2021 {\em Quantum Information Processing: Theory and Implementation} (Springer, Cham, 2021)
\bibitem{Schumacher}
Schumacher M and  Alber G 2023 Phys. Rev. A (submitted)
\bibitem{Sauer2}  Sauer A and Bern\'{a}d J Z  2022 Phys. Rev. A {\bf 106} 032423
\bibitem{Smith}  Smith R L 1984 Oper. Res. {\bf 32} 1296
\bibitem{Lovasz}  Lovasz L and  Vempala S 2006 SIAM J. Comput. {\bf 35} 985
\bibitem{Peres}
 Peres A 1996 Phys. Rev. Lett. {\bf  77} 1413
\bibitem{Horodecki}
 Horodecki R, Horodecki P,  Horodecki M and Horodecki K 2009
Rev. Mod. Phys. {\bf 81} 865
\bibitem{Gitt1}
Gittsovich O and G\"uhne O 2010 Phys. Rev. A {\bf 81} 032333
\end{thebibliography}
\end{document}